\documentclass{article}

\usepackage{microtype}
\usepackage{subfigure}
\usepackage{booktabs} % for professional tables
\usepackage[most]{tcolorbox}
\usepackage{listings}
\usepackage{xcolor}

\lstdefinestyle{promptstyle}{
  basicstyle=\ttfamily\footnotesize,
  breaklines=true,
  breakatwhitespace=false,
  columns=fullflexible,
  keepspaces=true,
  showstringspaces=false
}% hyperref makes hyperlinks in the resulting PDF.
\usepackage{tabularx}
\usepackage{graphicx}
\usepackage{subcaption}
\usepackage{dirtytalk}
\usepackage[font={small}]{caption}
\usepackage[labelfont=bf]{caption}
\usepackage[hidelinks]{hyperref}

\usepackage[accepted]{icml2021}

\icmltitlerunning{One stout to rule them all: Reconciling artificial intelligence, data science and malted alcoholic beverages}

\begin{document}

\twocolumn[
\icmltitle{One stout to rule them all: Reconciling artificial intelligence, data science and malted alcoholic beverages}




\begin{icmlauthorlist}
\icmlauthor{Dmitrii Usynin}{tum}
\icmlauthor{Elena Shmakova}{mpi}
\icmlauthor{Michael Rheinberger}{mpi}
\end{icmlauthorlist}

\icmlaffiliation{mpi}{Research Department Cell and Virus Structure, Max Planck Institute of Biochemistry}
\icmlaffiliation{tum}{Institute for Artificial Intelligence in Medicine and Healthcare and Institute of Radiology, Technical University of Munich}

\icmlcorrespondingauthor{Dmitrii Usynin}{dmitrii.usynin@tum.de}
% \icmlcorrespondingauthor{Eee Pppp}{ep@eden.co.uk}

\icmlkeywords{Artificial intelligence, Large language models, Data analysis, Fermented malt-based beverages}

\vskip 0.3in
]

\printAffiliationsAndNotice{}  % leave blank if no need to mention equal contribution

\begin{abstract}
Beer is a phenomenal beverage.
It has previously shaped the history of many peoples, states and cultures.
The beauty of beer is its versatility. Starting from the original implementations that were murky or diluted, over time researchers found novel approaches to gradually develop beverages that are diverse, intense and are pleasant for the end user.
Recently, the industry came up with the so-called \textit{craft beers}, that often differ from the commercial beers in production volume (due to lower capacities of the craft beer producers) and tasting profile (often having more intense unusual flavours). 
However, while it is often relatively easy to judge if a particular commercial beer is likely to be enjoyable, the same cannot be said about craft beers, as there are far too many styles, implementations and ingredients involved in their production.
This creates a gap between the beverage producers and the consumers due to the inability of the former to judge the preferences and the consumption trends of the latter.
As a response to this challenge we present a novel collaborative beverage-related data collection and analysis framework - the Distributed Beverage Analysis (DBA).
The idea behind this study is to identify the common trends and support them by empirical evidence to better understand the needs of the consumers.
We empirically verify DBA at the biannual \textit{Kraft Bier Fest} conducted by Vienna Kraft brewery in (you guessed it) Vienna.
To showcase a need in such kind of analysis, we evaluate various large language models (LLMs) against our collaborative framework and confirm that many AI models cannot be reliably used to reason over the trends and patterns in the evolving world of craft beer.
\end{abstract}

\section{Introduction}
\label{sec:intro}
Beer is a beverage that has been available as early as $9000$ B.C. \cite{wang2021earlybeer} and has been consumed daily by people of all nationalities and social classes (despite later re-classification during the early Roman Empire \cite{strickland2007beerbarbarism}).
It was consumed both as the beverage that brings people joy \cite{wang2021earlybeer} as well as a beverage to supplement the working people with additional calories and quench their thirst \cite{neumann1994beercompensation} (it was previously commonly believed that beer was used as a replacement for water, but recent studies have disputed this claim \cite{sheffield2018alcoholwater}).
In the early days of humanity beer was made almost accidentally \cite{livescience2025beerorigin} and once we (the humanity) determined the (somewhat) specific scientific approach to making this beverage, the production has gained some notion of consistency.
Fermented malt-based beverages have since become easier to store, produce and, most importantly, customise to the taste of the consumer (which in many cases is the brewer themselves) \cite{hornsey2003history}.
The final point is essential for this work, as many brewmasters (those responsible for the direction and the production of beverages at individual breweries) do not follow the fundamental scientific principle of \say{just because one can does not mean one should}, resulting in a wide array of experimental styles which are a) not classified as beer in certain countries anymore (see Section \ref{sec:discussion}) or b) contain flavours which are certainly courageous, but are very different to the expectations of a typical consumer.
But as the industry is evolving, so do its individual components, including the consumer preferences.
It is, therefore, much harder for the producer to predict the success for many of their newer products, due to a surge in the number of styles, implementations and the consequent reactions of the consumers.
The problem can often be further exacerbated by the inability of a typical consumer to determine if a beverage tastes in a peculiar way due to its poor quality, deliberate design decision or a shift in production trends (i.e. some styles slowly becoming more prevalent).
As a result, we observe a clear need for a  framework which would unify the ever-changing trends in beer design, production and the consumer preferences.
We hypothesise that a promising direction towards that goal can be found in the domains of data science and data analytics.
However, in order to analyse the data and have a better understanding of these trends, we first need to identify how such data can be collected, processed and aggregated across a variety of different consumer types.
To alleviate this problem we propose a novel data aggregation framework, termed Distributed Beer Analysis (DBA), which is designed to permit a group of consumers to collaborate during the tasting events in order to construct representative tasting data that can be used for further analysis.
We then apply this framework to collect data during a popular craft beer tasting event KraftBierFest hosted in Vienna to create a representative dataset and study the aggregated results in order to identify the common patterns, preferences and beverage implementation choices that lead to higher consumer satisfaction across a diverse group of tasters.

There is, of course, a temptation for the reader to ask: \say{why do you need to put all this effort in, when I can just post a beverage list into my favourite LLM and ask it to summarise the trends and make personalised beverage recommendations?}
To respond to this question, we evaluate our framework against a number of popular LLMs and showcase that despite the recent advances in this field, these models are not yet able to perform these tasks with high utility, missing on many subtle (or sometimes very obvious) details that a human participant can easily spot.
As large-scale LLM evaluation and benchmarking is not the scope of this work, we limit ourselves to models that a typical laptop user would be able to run at a similar tasting event in real-time. 

The contributions of this work are as follows:
\begin{itemize}
    \item We propose a novel distributed beer tasting and analysis (both qualitative and quantitative) protocol, termed the Distributed Beer Analysis (DBA)
    \item We collect a new representative craft beer dataset and showcase the effectiveness of this protocol during the KraftBierFest tasting event 
    \item We analyse the results of the tasting procedure both quantitively and qualitatively and summarise the main trends, directions and consumer preferences, bridging the gap between the local producers and their target audience.
    \item Finally, we compare our collaborative obtained results against a number of popular language models to showcase that these are not yet capable of understanding the trends and patterns of the craft beer consumers.
\end{itemize}

\section{Background}
\label{sec:background}
To start our discussion off, we need to firmly agree on what \textit{craft beer} actually is.
As per the definition of \cite{brewersassociation_definition}, this is a beer produced on a scale of fewer than $6M$ barrels per year (this definition is US-specific, but similar, albeit scaled definitions can be found for other markets).
This beer production is best described as small, independent and innovative.
As per the definition, craft beer producers \say{interpret historic styles with unique twists and develop new styles that have no precedent}.
This experimental mindset allows the craft beer producers to create novel experimental styles that better represent their innovative mindset.
An additional benefit of craft beer is that while the projected profits are likely smaller than compared to a large commercial producers, the price of error is also significantly lower, allowing for a higher degree of creativity during the process.

Now that we established what craft beer is, we briefly discuss prior works that study consumer trends, preferences and view on craft beer.
As described in \cite{aquilani2015beer}, consumption habits for craft beers can be loosely split into beer-related and purchase-related.
We have already touched on craft beers having unique tasting profiles compared to their commercial counterparts, meaning that consumer preference patterns are strongly affected by the taste of the beverage they consume.
However, the consumption trends are also heavily influenced by the macroeconomics of craft beer production: since small producers often do not enjoy economies of scale, the price per beverage can be significantly higher \cite{craftypint_costs}.
While no doubt very important, the macroeconomic aspect of craft beer falls outside of the scope of our work: all samples in our study had the same cost of $0$ EUR, meaning that we could not derive any new insights from participants' spending habits.
Other similar studies have previously analysed and quantified consumer preferences and habits across various geographies ranging from Brazil \cite{da2018sensory} to Italy \cite{garavaglia2020craft} and Mexico \cite{gomez2016craft}.
It is, however, difficult to objectively judge how well the results of these studies scale across different geographies.
Additionally, these studies primarily concentrate on analysis of existing data sources and do not attempt to obtain or aggregate new datasets.
To the best of our knowledge, our study is the first empirical attempt to quantify these trends across multiple geographies, conveniently represented by various producers at the same location, drawing crowds of beer enjoyers from all over the European continent.

\section{Methods}
\label{sec:methods}
To generate the ranking data we assembled a panel of $3$ expert judges (as well as $5$ amateur judges, who we list in the acknowledgements section, but whose scores were not included in the results section, as they did not bother writing these down).
The scores lie between $1$ and $5$ and we share both the normalised and the unnormalised version of the rankings.

\begin{figure*}[t]
    \centering
    \includegraphics[width=\textwidth]{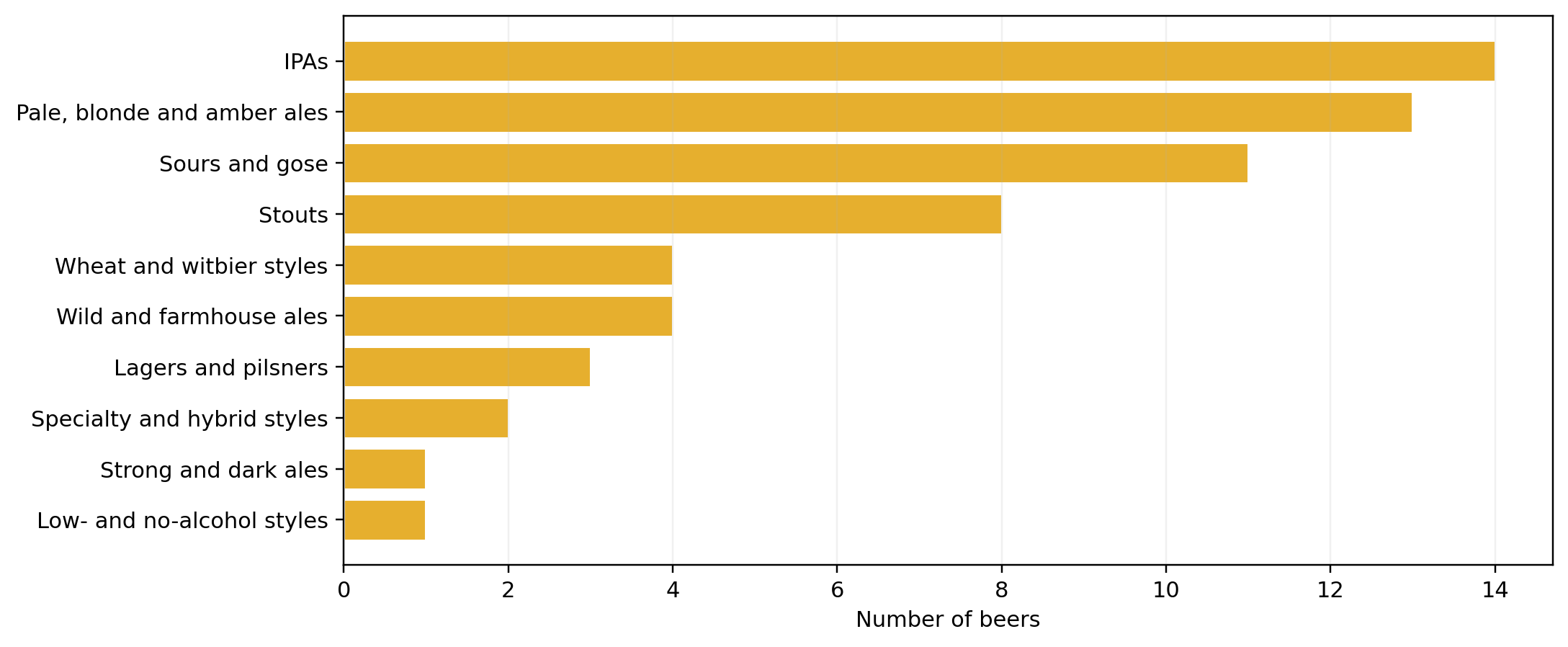}
    \caption{Overview of the style families present in the dataset. The dataset is dominated by ale-based styles and sours.}
    \label{fig:style_counts}
\end{figure*}

\subsection{Per-round data, client and leader selection}
The setup of the experiment is very similar to the one of the hierarchical on-device federated learning \cite{liu2020hierarchicalfl}.
Each member of the expert jury acts as a rotating central server.
The leader selection is probabilistic - judges A and C have an identical probability of $0.1$ to become the leader for the next round, whereas judge B has a probability of $0.8$.
Once the leader announces the start of the new round, the federation (those that are currently able to participate) selects (obtains from a producer) a single sample without replacement, analyses the current data item and shares both the resulting review as well as the beverage with other participants.
Unlike the standard FL protocol, DBA allows (and somewhat encourages) freeloaders who do not contribute to the procurement of data items, but share the final review.
In practice this often means that at each round the number of participants is smaller than the size of the federation minus the leader (which is expected for on-device FL), but the number of datapoints is also smaller than the number of participants (which is typically not the case in any modes of FL to the best of our knowledge).
It is important to note that due to the nature of the protocol (i.e. that it is human-in-the-loop by design), the availability of participants is biased with respect to time of day.
During time periods such as lunchtime or dinnertime, the availability is reduced and, hence, instead of reporting the biased results of the protocol (as there would be fewer, likely more imbalanced reviews), we omit these altogether.

\subsection{Numerical ranking of beverages}
Each of the expert judges (protocol leaders) were tasked with producing both an absolute numerical score for each beverage as well as the qualitative assessment of how they arrived at that score.
The latter normally described characteristics of the beverage and outlined the reasons behind the final rankings.
For the remaining five participants, only the numerical scores were provided and they were primarily used for calibration, rather than as additional data points.
It is important to note that the original scores were unnormalised, and for some judges that meant that they were frequently tapping into the long tails of the distribution of the scores (e.g. multiple $1$s were reported by judge B).
To account for this issue we performed standard score normalisation so that the scores range between $0$ and $1$ instead.
It is of note that this has not significantly changed the aggregated results, but allows for a better degree of interpretability for the reader.

\subsection{Qualitative assessment of the items}
Finally, as we have already alluded to before, the judges were asked to produce qualitative assessments behind their scores.
As the task was open-ended, this meant that the assessment included some subjective statements about the intensity beverage (e.g. \say{strong mango flavour} by judge C) as well as the subjective perception of some of its characteristics.
A notable example of this is a comment made by judge B: \say{tastes of sweaty armpits}, which typically indicates presence of various off-flavours that could have originated during improper storage or transportation of the beverage \cite{brewersassociation_isovaleric}.
It is, however, of note that the origin of this flavour does not concern the end user, meaning that while the original producer is often not responsible for such flaws, they are ultimately bearing the cost in terms of poor reviews, reduced consumer trust and future revenue losses.

\section{Results}
\label{sec:results}
In this Section we showcase the trends, patterns and numerical results of the expert rankings.

\subsection{Dataset description}
We start this section with a brief overview of the samples used in this work.
The dataset used in this study consists of approximately $60$ malt-based beverages, ranging in alcohol by volume (ABV) from $0.5\%$ (limit at which the beverage is considered to be alcohol-free) to $12.5\%$ (for some imperial stouts).
The majority of samples have $3$ reviews, but each sample is guaranteed to have at least $2$ reviews.
We discuss the nature of the qualitative reviews in Section \ref{sec:qualitative}.

\subsection{Common trends}
From the brief overview of our data we can easily observe the expected main trend: craft beer is almost universally linked to Pale Ale-derived styles and Indian Pale Ales (IPAs) in particular.
Another (somewhat expected) trend is the abundance of experimental styles, many of which are specific to the European region.
These include sour styles (native to Belgium), stouts (native to the UK and Northern Europe) and wheat-derived beers (naive to the DACH region as well as Belgium).
So we observe that while there is a healthy mix of styles present in the dataset, lager-based styles are significantly less present, likely due to the event prioritising novel, experimental beverages.
We hypothesise that these results demonstrate our first preliminary finding: producers are expecting high level of interest in IPAs, sour ales and stouts.

Next we cover the relative strength of beverages, using a scale inspired by WSET beer guidelines for labelling based on ABV (low is $<4.5$, medium is $4.6-6.5$, high is $6.6-9.0$ and very high is $>9.0$).
In terms of relative strength, we see a range of ABVs, with the majority in medium region.
Again, this is somewhat expected, given that producers want their beers to be unique, but approachable for those new to craft beers.
The distribution of very strong beers does not come with many surprises either: all of these are imperial stouts of some form.
One notable exception to that were the strengthened forms of IPA (double IPA) and the grape ale (where grapes are used as past of the fermentation process).
The second preliminary finding we identify here: beer strengths demonstrate a lot of diversity to attract a variety of tasters, with the majority being of medium strength.
This caters to the majority of consumers by not overwhelming them with unnecessarily strong beverages, while allowing those interested in more heavier and full-bodied styles to explore those towards the end of the event.

\begin{figure}[h!]
    \centering
    \includegraphics[width=\columnwidth]{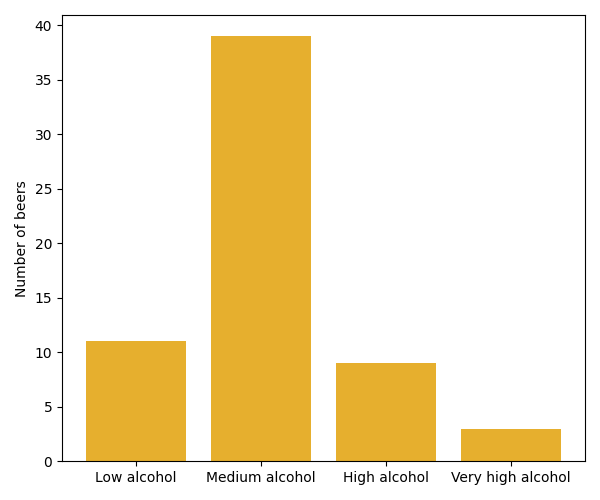}
    \caption{Overview of beverages separated by ABV. The majority of beverage have medium ABVs.}
    \label{fig:abv_chart}
\end{figure}

\subsection{Styles and producers}
In this study, we evaluated $19$ producers, each allowed to present up to $4$ beverages.
As part of our data aggregation, we have collected data for each beverage available from each producer with an exception for the hosts of the tasting (Vienna Kraft), since the judges have already sampled their products at a similar event before.
While there is a lot of variation in the exact styles that each producer has presented, for the purposes of interpretability and data normalisation we have grouped these into 
$10$ overall style buckets.
The overview can be found in Figure \ref{fig:style_counts}.
In general, we note that most producers demonstrated many diverse styles (e.g. Abodee with $3$ samples and $3$ unique styles), but some tried to play to their strengths and present beers of a specific style to showcase their experience in these (e.g. Furst Wiacek with $3$ IPA-based beers).
In terms the producers themselves, we see that there is a lot of diversity in their countries of origin: Germany, Austria, Hungary, Scotland and England.
The somewhat unexpected trend we observe is that regardless of the country of origin, producers do not stick to their \say{native} styles, such as the British bitter for any of the UK-based producers.
Our hypothesis is that this is most likely due to a combination of the desire to cater to an international crowd (by using more well-adopted craft beer styles such as IPA) as well as the difficulty of storage and transportation for some of the native styles (using the same example of bitters - ales have a notoriously short shelf-life, meaning that transportation could result in spoilage \cite{briggs2004brewing}).

\subsection{Quantitative results and analysis}
\begin{figure}[h!]
    \centering
    \includegraphics[width=\columnwidth]{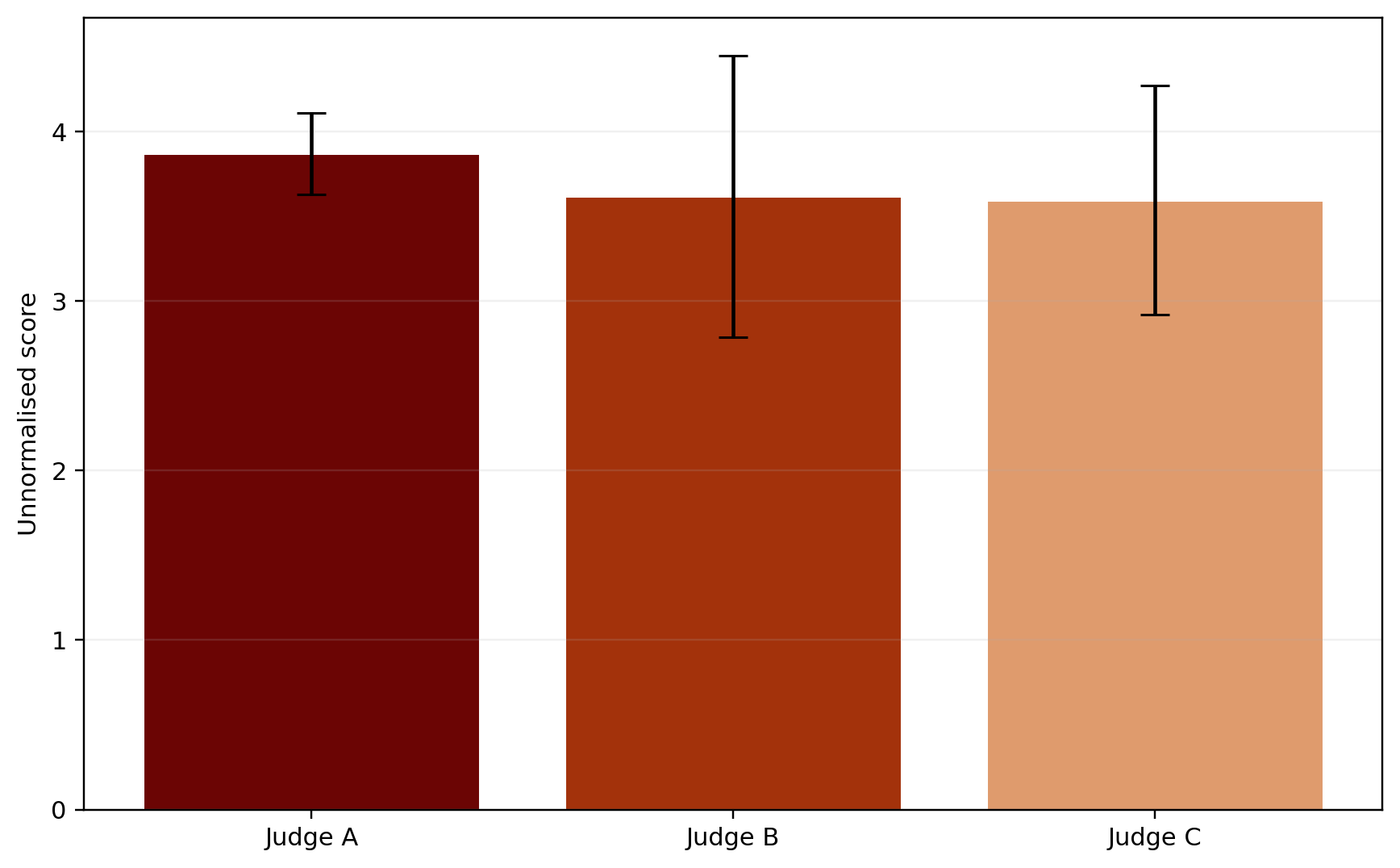}
    \caption{Mean and standard deviation for each judge, unnormalised scores. Judge A has, in general, been more conservative with their score distribution compared to other judges who were more likely to produce more extreme scores.}
    \label{fig:judge_mean}
\end{figure}

\begin{figure}[h!]
    \centering
    \includegraphics[width=\columnwidth]{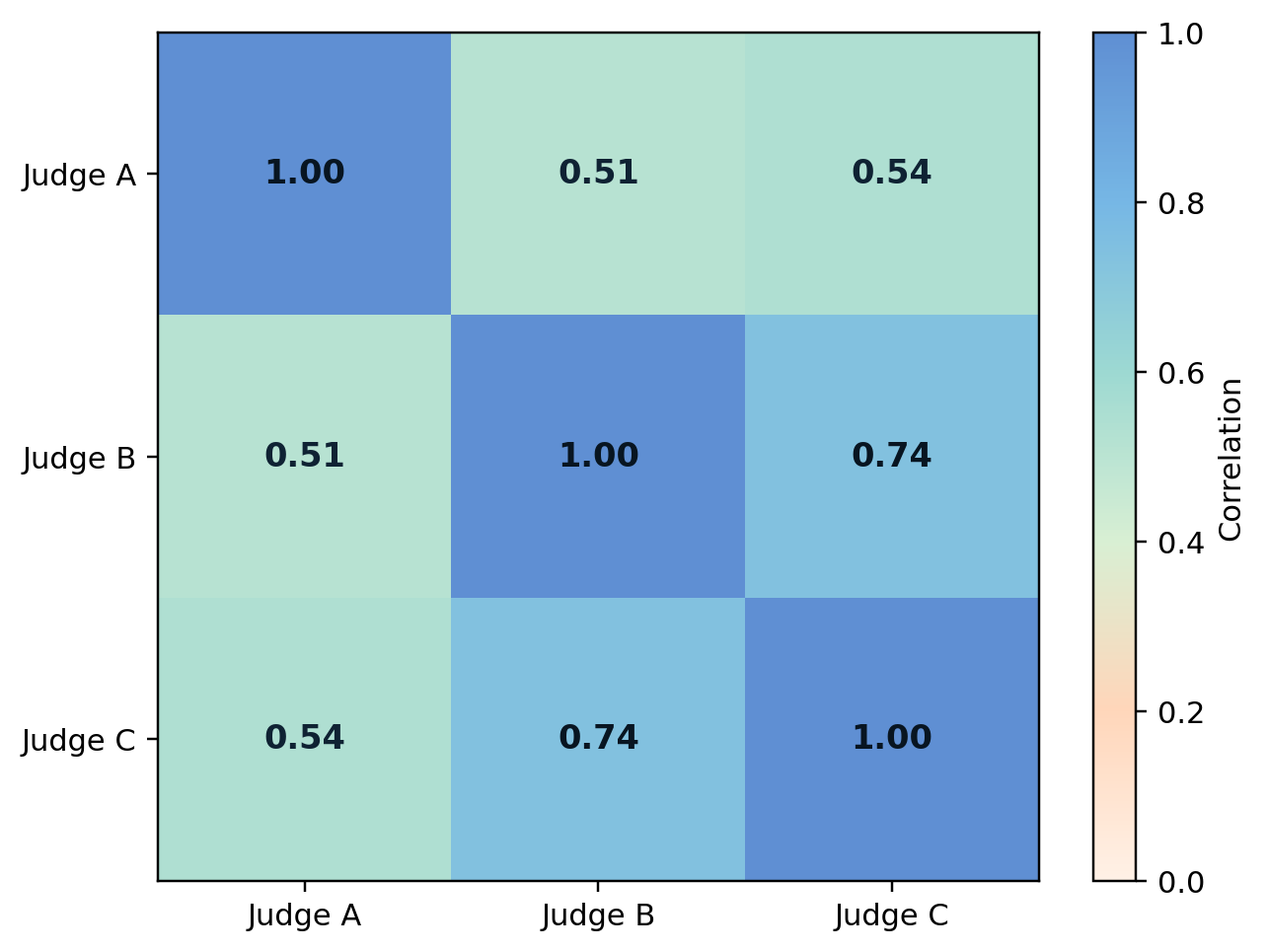}
    \caption{Agreement between the judges. Judge B and C have a much higher overlap in rankings compared to judge A.}
    \label{fig:agreement}
\end{figure}

Now that we have discussed the data in more detail, it is time we analyse how the expert judges have ranked the beverages during the event.
We begin this section with an overall assessment (Figures \ref{fig:judge_mean}), which shows that the mean unnormalised scores lie between $3.5$ and $4.0$ for all judges and that despite having similar means, judges have often disagreed on individual scores, highlighting the importance of a diverse set of tasters.
The agreement chart between judges can be found in Figure \ref{fig:agreement}.
We include the complete ranking table (unnormalised to showcase how varies were the responses of individual judges) in the Appendix.
We used the characters from the AppleTV show \say{Shape Island} to represent the individual judges on the score chart as these seems to correlate with the personalities of the judges relatively well.

\begin{figure*}[h!]
    \centering
    \includegraphics[width=\textwidth]{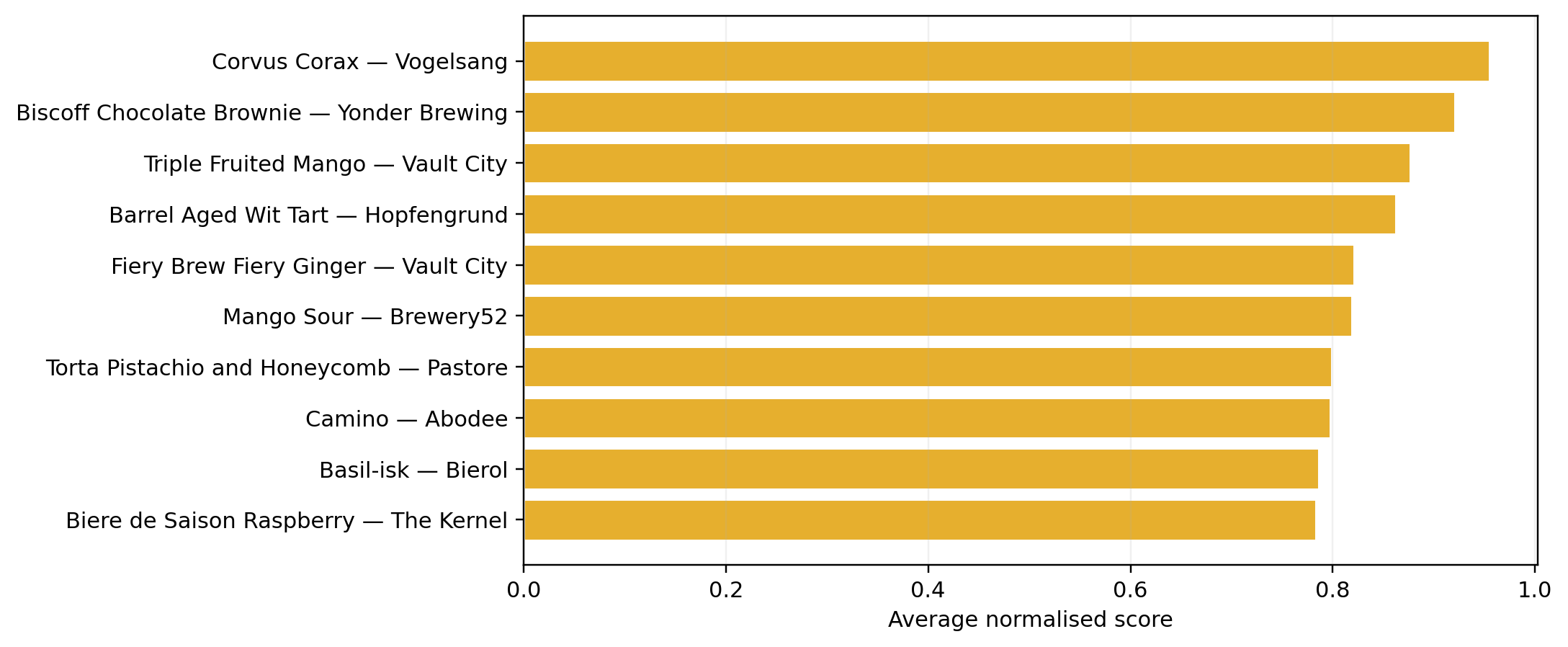}
    \caption{Top-$10$ beverages, normalised. The majority of beverages are stouts, sours or wild ales.}
    \label{fig:top10}
\end{figure*}

\begin{figure*}[h!]
    \centering
    \includegraphics[width=\textwidth]{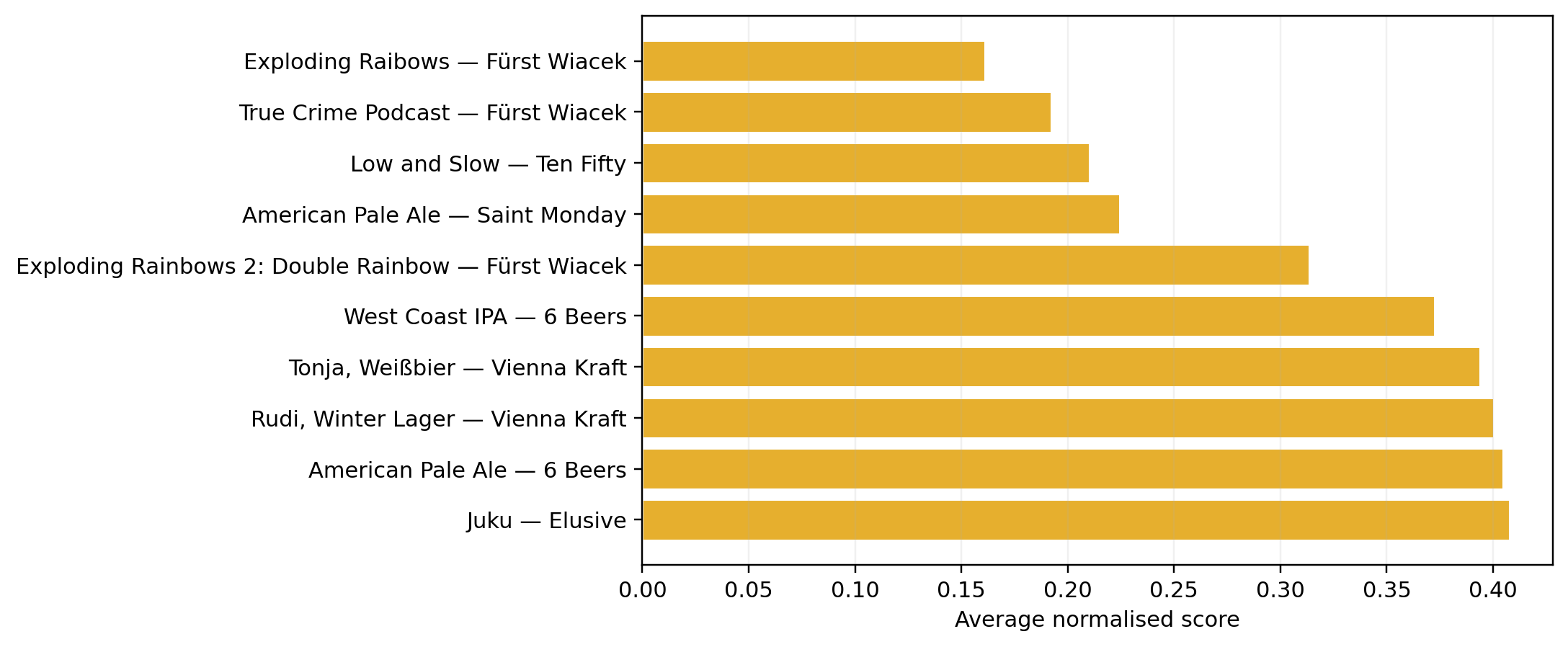}
    \caption{Bottom-$10$ beverages, normalised. The majority of beverages here are lager-based or pale ale-based.}
    \label{fig:bottom10}
\end{figure*}

We present the overall (normalised) scores in Figures \ref{fig:top10} and \ref{fig:bottom10} for the top-$10$ and the bottom-$10$ tasted beverages respectively. 
To better understand the effects of individual preferences, we additionally include the rankings per each style made by each judge in Figure \ref{fig:judges_boxplot}.

\begin{figure*}[h!]
    \centering
    \includegraphics[width=\textwidth]{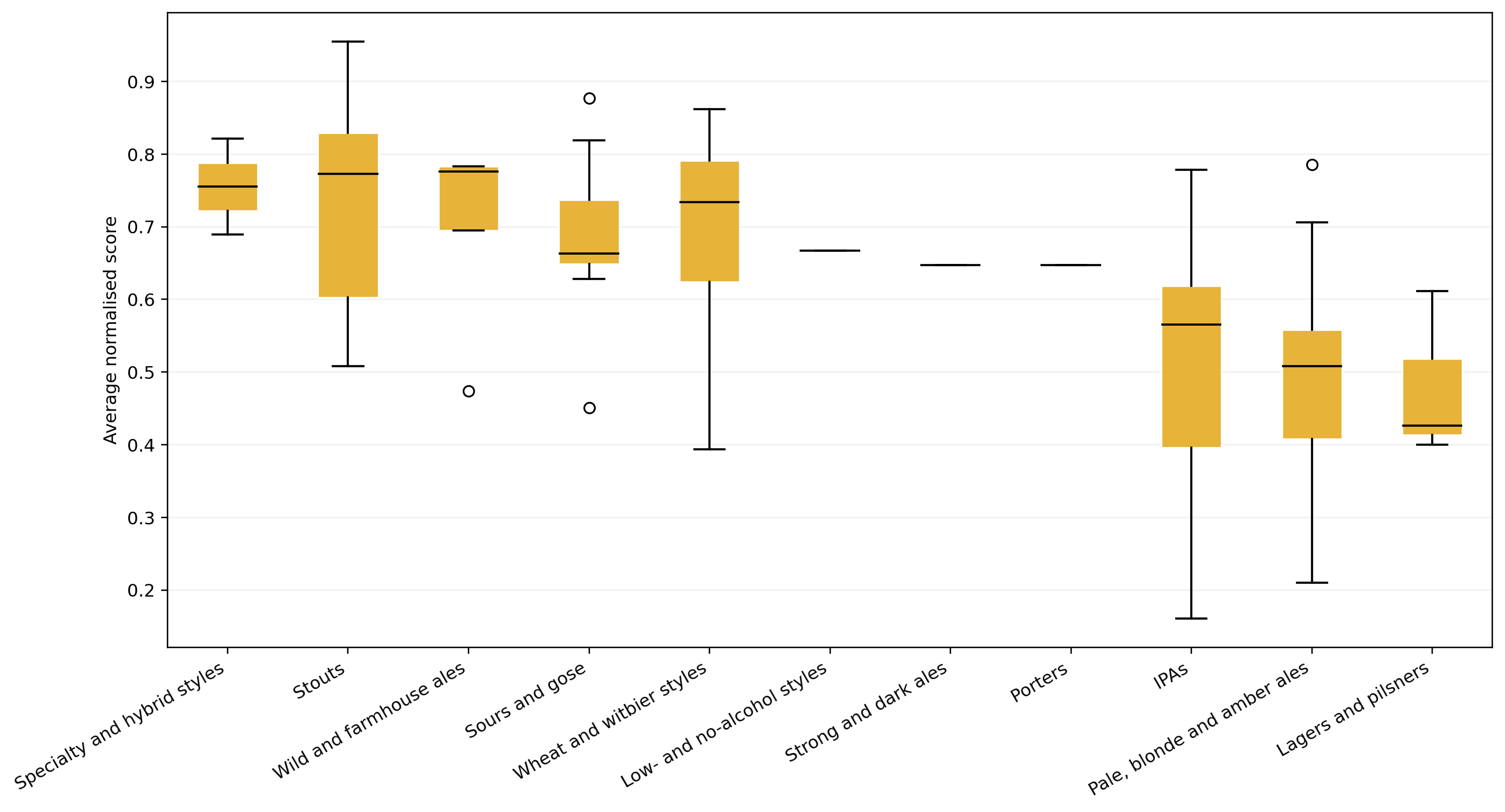}
    \caption{Distribution of aggregated scores across different styles. Some styles were underrepresented at the event, resulting in a single score per style.}
    \label{fig:per_style_normalised}
\end{figure*}

Next we discuss how individual styles have performed (in contrast to individual beverages). 
We demonstrate our results in Figure \ref{fig:per_style_normalised}.
The results have to be interpreted with care here. 

\subsubsection{Pale Ale-based styles}
First of all, as we have previously discussed, IPAs and Pale Ale-based styles were disproportionally overrepresented at the event, meaning that the majority of producers likely considered them to be more popular among the consumer population (or to be safer choices, which are relatively easier to \say{get right}).
So we are in a position where such styles are a) common and b) relatively similar to each other in flavour profile (as per judge C regarding one of such beverages: \say{quite unexciting}).
As a result, the majority of judges agreed that due to a lack of variation, overrepresentation of the styles and general similarity to the more common commercial brands, the overall scores for these styles were lower.
One additional important note that we cover in the next subsection is the fact that such styles are typically slightly more challenging to transfer and store.
The bright fruit notes in most ales have the tendency to be muted when stored in incorrect conditions for too long with a high risk of being converted in off-flavours which are much more pronounced compared to most other non hop- or fruit-forward styles.
Therefore while the lower scores for such styles are expected, there is a large list of factors contributing to these, not necessarily indicative of the quality of the individual beverages.

\subsubsection{Wild and sour ales}
In contrast, sours and spontaneously fermented beers are were in the spotlights at this event: these beers are relatively difficult to procure in supermarkets, have very intense distinct flavours and are unlikely to be similar to each other due to the spontaneous fermentation process.
As a result, these styles enjoyed higher scores on average, again, regardless of the quality of an individual beverage.
In general, many of these beers could be perceived as close to fermented fruit juices (or even ciders), making them slightly more appealing to the non-expert tasters.
While sensitivity to high acidity can be a detriment to the enjoyment of these beers (judges A  and C have previously declared this as a potential source of bias), their unique flavour profiles are unmatched and these often scored highly simply for how unusual they were.
An example of that is a remark from judge C on the quality of the bergamot sour: \say{smells very funky, tastes good but not as funky as the scent makes you expect, quite sour}, giving it a score of $3.8$.
This remark demonstrates that despite some off-putting characteristics, such styles are simply \say{too interesting} not to try and they often end up being more positive than originally anticipated.
The fact that there is no strict definition to what a sour beer is; or what is its desired flavour profile have also contributed to their relatively large presence at the event (as indicated in Figure \ref{fig:style_counts}).

\subsubsection{Dark beers and stouts}
We want to also highlight the more full-bodied darker beers encountered during the event.
One of the most commonly known examples of such beers are dry stouts (e.g. Guinness) and these are often difficult entry-level beers, due to complex burnt flavours and pronounced bitterness. 
As a result, enjoyment of darker beers is often described as an \say{acquired} taste, which comes over time and this makes many such beers unattractive to inexperienced tasters.
Therefore, for many producers stouts are an important element of their portfolio, but is likely not their main revenue generator and often does not attract as many customers as more easy-drinking ale-based styles.
However, during events such as the KraftBierFest, where the majority of the consumer population consists of people who have been trying and enjoying many different styles of beer for a long enough time to know about the existence of such events, the picture is rather different.
There was a variety of dark beer styles ranging from the classic dry stouts, to black IPAs to imperial stouts, all of which focus on completely different parts of the flavour spectrum.
For the final scores, we observe various contributing factors: a) these beers are not frequently encountered commercially, b) the consumer base of the event was likely experienced enough to enjoy the nuances of the style and c) the fact that many of these are expensive to acquire (and were already included in the ticket price of the event) - have all contributed to their high scores across all $3$ judges.
As far the quantitative results show, stouts were consistently ranked as the most highly-rated style.
This was in particular due to the inclusion of pastry stouts and imperial stouts: two distinctly different styles, united by their relatively high ABV (giving them additional body and creating a warmer aftertaste) and high sweetness, masking away the bitter or burnt flavours typically associated with stouts.
We note that it was not a coincidence that out of $3$ top-scoring beverages, $2$ were stouts.
As we have already described above, this is due to a combination of their unique \say{positive flavours} (chocolate, caramel, pastry to name a few), well masked \say{negative flavours} (bitterness, burnt bread, harshness) and difficulty of obtaining these beers elsewhere (due to high price and low availability).

\subsubsection{Lager-based styles}
Finally we wanted to outline a short overview of the remaining styles, most of which are malt-driven lagers.
In general, one would find that these beers generate the most revenue worldwide \cite{brewingscience_lager}.
They are often very quaffable (i.e. easy to drink in large volumes), refreshing, easy to obtain and are not particularly strong, allowing the consumer to enjoy multiple portions in a single sitting.
The downside of that, however, is that they are simply not \say{interesting} enough to be compared against experimental styles at such events.
And while some implementations can be executed to perfection (primarily because they are really well-balanced and demonstrate the perfect representation of the desired flavours), their availability and lack of experimental profile, typically result in lower scores at such events.
We cannot stress this enough: the main driving factor behind this is not their poor quality or misrepresentation of the general consumer trends, but simply the difficulty of outcompeting the already experienced consumer base at craft beer tasting events, where the scoring trends are inverted.
Consumers tend to often give higher scores to less-balanced and highly unusual varieties in contrast to already widely-available lager-based styles.

\subsection{Tasting notes and qualitative insights}
\label{sec:qualitative}
While the numerical results give us a good overview of trends, patterns and consumer preferences, as we have already discussed above: it is not always possible to establish the specific reason behind each individual score.
To better understand why certain seemingly identical taste profiles can produce distinctly different rankings, we study the notes the judges made during the event.

\subsubsection{Fruits, berries and other atypical ingredients}
Beer is a beverage typically made out of hops and malts of various kinds using yeast during the fermentation process.
The reason this process works is due to the conversion of sugars contained in the malts into ethanol and carbon dioxide.
What is often overlooked is that malts are not the only ingredients that can contain natural sugars that can be fermented: many fruits and berries are also full of sugars that can be used during the beer fermentation process.
As a result many fruit co-fermented beers have made their way into the festival, the majority of which have scored relatively highly regardless of the preferences of the judges.
Judge B, for instance, has awarded the Mango Sour by Brewery52 with a score of $4.8$ and a comment: \say{smells like a real mango and tastes like one}.
In general, as we have already seen in our quantitative assessment, the majority of \say{unusual} flavour combinations scored relatively highly, but the fruit-dominated varieties had the largest difference between what the beer \textit{smells like} and what it ultimately \textit{tastes} like.
Fruit-forward beers can come in a variety of styles such as saison (e.g. Raspberry Saison by the Kernel) or the aforementioned sours. 
The important point here is much more on the presentation and marketing side than it is on the flavour profile of the beer.
We found that beverages which had a large disparity between the nosefeel and the mouthfeel scored significantly lower compared to those where the expectation matched the reality.
Prime example of that is the Blackberry and Cherry Granita by the Pastore brewery, that was awarded $2.7$ by judge C with a comment: \say{too sour, feels weird in the mouth afterwards, smells very different than it tastes}.
Despite the fact that overall, this beverage has scored relatively well compared to the rest of the dataset, it was not the flavour itself, nor its intensity that resulted in a lower score, but rather the problem of expectations.
There She Gose by Bierol had the exact same problem: judge A awarded it with $3.8$ and a comment \say{real bergamot flavour on the nose, normal gose on the palette}, indicating that the mismatch in expectations is the leading cause for a lower score compared to other beers of the same type.
What unites these two examples?
The style: both of these are sours.
We discuss the issue of flavour balance in the section below, but here we note that while the fruit-forward beers have scored higher than most hop-forward styles on average, they also suffered from a lot of divisions between the judges (particularly in the unnormalised version of the results seen in the Appendix \ref{fig:divisive_beers}).
As a result, we hypothesis that the fruits should be well-supported by the main flavour profile of the beers.
If we are expecting raspberry to be slightly sour, then sours and saisons are the right combination.
But if we are expecting a strong mango profile, it might be better suited for a hazy IPA instead.

\subsubsection{Flavour balance}
We have already discussed the unusually-flavoured beers at length in the previous sections.
What we have not explicitly touched upon yet is what kind of additional flavours are these perceived to be, and how does that affect the scores of a beverage.
To be a bit less cryptic on this, we will study a concrete example: the Vault City Brewery presented $2$ very different styles of sours - the Triple Fruited Mango and the Sour of Scotland.
The former has attracted numerous praises from all three judges (judge A: \say{best mango so far}, judge B: \say{real mango here}, judge C: \say{tastes and smells like mango Schorle}, which in this case is a strong positive signal).
The wording here speaks for itself: all judges praise not just the high intensity of the fruit flavour, but the fact that it tastes like a \textbf{real} mango.
In contrast, the Sour of Scotland has received relatively muted responses with judge A stating that it tastes \say{too artificial}, reflected in significantly lower scores.
Here we see that the score is reflecting not just the unusual flavour itself, but also how well-balanced it is by the rest of the beverage.
We see very similar results across all other \say{unusual} styles, where none of the ones with comments mentioning \textit{artificial flavours} scored higher than those that mentioning \textit{real flavours}.
This is pretty indicative of the fact that balance is one key factor which is often difficult to get right in a craft beer which relies on additional unusual flavours.
One conclusion we can draw from this discussion is that while the mere presence of a new, interesting flavour can severely increase its score, its over-the-top intensity can have the opposite effect.
The craft beer community does, indeed, appreciate novel ingredient combinations, but not at the expense of the flavour balance.
% TODO: talk about the latent features of the beers, where the taster thinks they like belgiums, but they really like a property of belgiums that is present elsewhere
\begin{table*}[t]
\centering
\small
\begin{tabular}{lccccc}
\toprule
Model & Mean rating & Mean percentile & Hit@5 & nDCG@5 & Coverage \\
\midrule
Phi-4         & 4.063 & 0.718 & 0.333 & 0.844 & 1.000 \\
Ministral 3B  & 3.967 & 0.605 & 0.133 & 0.743 & 1.000 \\
Qwen2.5 3B    & 3.920 & 0.559 & 0.133 & 0.852 & 1.000 \\
Qwen3 4B      & 3.890 & 0.581 & 0.133 & 0.774 & 0.867 \\
Gemma 3 4B    & 3.820 & 0.560 & 0.200 & 0.827 & 1.000 \\
Llama 3.2 3B  & 3.807 & 0.498 & 0.200 & 0.814 & 1.000 \\
\bottomrule
\end{tabular}
\caption{Recommendation quality of the evaluated language models against the participants' empirical beer ratings. Mean percentile is computed within each participant's own ranking, where higher is better. Coverage below 1 indicates invalid or missing recommendations. Note: the rating is unnormalised.}
\label{tab:llm_beer_results}
\end{table*}

\subsection{LLM-powered trend and preference predictions}
While the authors have thoroughly enjoyed participating in such study, a fundamental question we posed in the introduction remains unanswered: \say{what is the point of trying to to come up with all of this, when one can just ask a language model?}
And this question is not without its merit: in the age of \say{everything is a computer} \cite{trump_everything_computer_2025}, it is often far easier to treat large language models as a compact source of insights, which would respond to your queries in natural language in a digestible and highly customisable format.
We argue, however, that while LLMs are very good at representing general statistically probable events, craft beer tastings represent the exact opposite.
These often prioritise unpredictability, uniqueness and combinations which simply should not work, but somehow do.  
To demonstrate this point we evaluated the effectiveness of publicly available LLMs at predicting: first the most preferable beers the judges would choose and second the general trends and patterns of craft beer consumption.
As we have briefly outlined in the introduction, we mainly evaluate the models which can easily be run on consumer-grade hardware in real-time e.g. by a participant at a similar tasting event.
We selected the following popular models for evaluation: Qwen$2.5$-$3$b, Qwen$3$-$4$b, ministral-$3$-$3$b, Gemma$3$-$4b$ and Llama$3.2$-$3b$.
We additionally include Phi$4$ as a slightly larger model to analyse if there is a correlation between the prediction accuracy and the model size.
We run all of these models locally on a MacBook Pro with an M$2$ chip and $16$Gb of memory.
The prompt we used can be found in the Appendix and it consists of short personal preferences of the judges as well as the list of beverages they have sampled.
The idea behind this experiment is to a) determine if popular language models could suggest suitable beverages based on the user profile and b) if they have a firm grasp over trends and patterns of craft beer production and consumption.
We report our results in Figure \ref{tab:llm_beer_results} as well as how well these correspond to the ground truths (i.e. expert opinions of the judges).
What we observe here is that a) the majority of the models are falling back to generic craft beer related trends (e.g. Qwen$2.5$ claims increased popularity of Hazy IPAs, of which there are only $2$ in the dataset), b) they seem to suggest inappropriate beverages (e.g. Brett Pils is still a Pils, making it a poor choice for judge A) and c) the model seem to arbitrarily ignore certain constraints (e.g. suggesting the same beverage twice or refusing to suggest $5$ in total because there are no suitable ones in the list).
In general we find most models to be sub-optimal, as there is very little overlap with what the judges have rated as their top choices.
In some cases the model has ignored the preferences of the individual altogether and suggested generic, well-known craft beer styles unsuitable for this specific consumer (e.g. Gemma$3$ not suggesting a single mango-flavoured drink for judge C).
Overall, while larger models such as Phi$4$ have produced much better personalised predictions, none of the models were able to accurately match the profile of the individual to the beverages they have selected as their top choices.

\subsubsection{Personal preferences and latent features}
What we do find interesting here (and leave a detailed evaluation of this as future work) is that in some cases, the models (e.g. Qwen models or ministral) have correctly picked up on the \say{shallow} features of a beverage (e.g. the fact that Saison is often classified as a Belgian style or the fact that Brett yeast is typically used in Belgian beer styles), but was incorrect in the final suggestion.
What does this tell us about the effectiveness of LLMs at this task? 
Precisely what we have hypothesised before: without doing a large-scale empirical evaluation, it is simply not possible to make informed decisions in a domain where the description of a data point may not match the content of the same data point.
To be more specific: as the self-description suggests, judge A prefers \say{strong, but not overly strong beers}, which goes against their own final rankings: their top selection included $3$ beers with ABVs well above $9.0$.
The fundamental self-description is not incorrect: the judge does indeed prefer Belgian strong beers (such as triples and dubbels, both of which are high not very high in alcohol), but ultimately ranked a completely different style the highest.
This, however, is significantly affected by the quality of the implementation of the beverage, where the imperial stouts in this event did not have the undesired flavours associated with extremely high ABVs associated with this style.
As a result, we, once again, observe the phenomena we have previously seen in a variety of other tasks: the model will perform at best as well as a human would when presented with incomplete information.
In this competition, the judges had their prior assumptions on which styles they would enjoy based on their previous experience.
However, craft beers are by design more experimental and varied based on the specific implementation, making such priors significantly less usable.
The pattern that we do see, however, is that both the judges and the LLMs (the more performant ones anyway) make predictions based not on the beverage itself, but on the characteristics that pertain to a specific style the beer claims to be following.
If we analyse the top scorers (using judge A as an example), we see that there is a so-called \textit{latent} representation of their preferences in some of the selected beverages.
What the judge wanted was a \say{malty, strong} beer, and while typically imperial stouts would violate the constraint of \say{not overly strong}, in this specific instance the quality of the implementation has hidden away some of these undesired characteristics to the point where they are not noticeable, while keeping the latent representation of \say{malty} and \say{strong}.
This important distinction between what a craft beer \say{tastes like} and \say{should taste like} is precisely what makes this field so exciting, but also what makes evaluation based solely on the priors (which is the only information the LLMs have access to) miss out on subtle details and ultimately result in unsuitable predictions.

\begin{figure*}[h!]
    \centering
    \includegraphics[width=\textwidth]{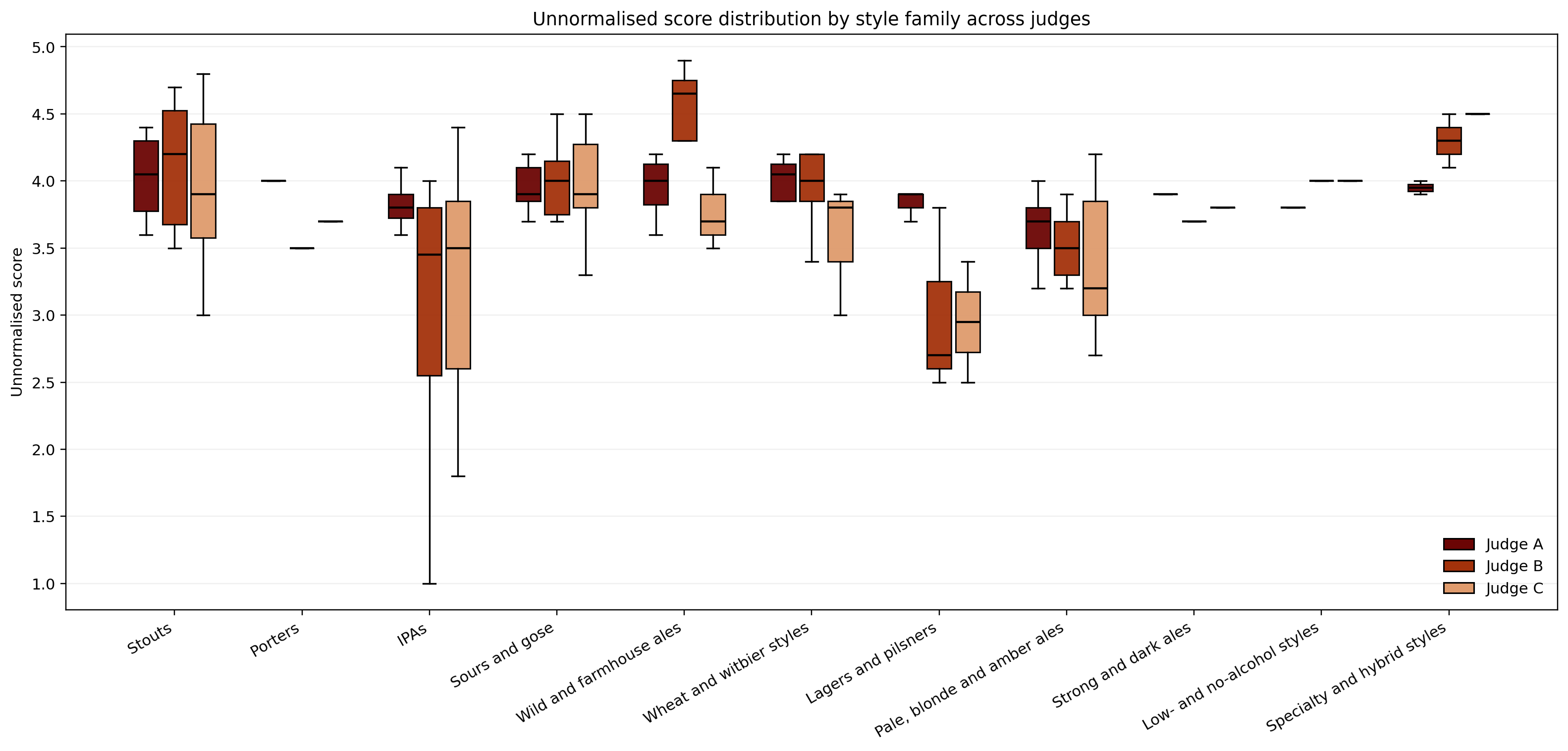}
    \caption{Scores spread of each judge per each style. Here we demonstrate why normalisation is required a) per value (as some the judges have distinctly different score ranges) and b) across multiple judges (individual biases are very strong, leading to skewed results).}
    \label{fig:judges_boxplot}
\end{figure*}

\subsubsection{Overall trend analysis}
While we expected the models to struggle with personalised predictions, what LLMs are known to be good at is pattern analysis \cite{ma2023insightpilot, jansen2025leveraging}.
However, as we have discovered above - this domain is far too challenging to reason over based on the priors alone.
In some cases (e.g. for Llama), it came up with generic trends not supported by the input data it was presented with.
Other models (e.g. Gemma or Qwen$2.5$) reported a mixture of generic and data-supported results, again, claiming a rise in popularity of hazy IPAs and low ABV beverages. 
Some models (e.g. Qwen$3$) have repeatedly failed to respond in the expected format, attempting to instead reason over the trends of the $3$ judge profiles.
Surprisingly, the model that previously performed the worst on recommendations side, has given the most appropriate response on the trend analysis (ministral).
In short: all models, including the larger Phi$4$ have reported either completely baseless trends or trends which included some factually correct information (e.g. the emergence of many unique variations of sours), mixed with commonly expected insights, which did not apply to our data.
Once again, we showcase that such \say{companion-LLMs} are not yet powerful enough to be used to analyse production trends or consumer preferences in such settings.
We share the resulting generations in the Appendix.

\subsection{Collaborative rankings vs individual rankings}
As expected, any individual assessing craft beers on a numerical scale is prone to biases and overpersonalisation. 
To demonstrate that we provide numerical comparisons between different styles across all judge scorecards.
As seen in Figure \ref{fig:judges_boxplot} the personal opinions of the judges very clearly represent their personal preferences (e.g. highest scorers being stouts for judge A and sours for judge B), but the overall trends remain consistent.
The highest scores were awarded to beers that a) had unique flavour profiles (which often happen to be sours or wild ales), b) were well-balanced (e.g. most imperial stouts at the event) and c) were hard to obtain in regular stores.
Therefore, we, once again, observe that diversity of preferences, experiences and backgrounds plays a critical role in estimating consumer preference trends.

\section{Discussion}
\label{sec:discussion}
\subsection{Trends in beer styles}
Overall we notice that the majority of styles present in the dataset are ale-based beverages (in contrast to lager-based styles commonly present in commercial production).
Subsequently we also note that pale ale-based styles were not ranked particularly highly by any of the judges, as these styles did not show a lot of flavour diversity.
As judge A described one of the West Coast IPA samples: \say{default WCIPA, hoppy, nothing special}, demonstrating that while these styles are almost always \say{more interesting} than their commercial counterparts, the \textit{craft} aspect in their design has been commonly adopted by many large-scale productions, rendering them overall a lot more uninspiring compared to other styles.
However, the exact opposite can be seen in less widely adopted \textit{wild} styles, which a) show the largest degree of variation, b) are unusual in taste, and, hence, more prone to extreme scores and c) can end up different to what the producer had in mind.
Some of these styles rely on wild yeast, meaning that each batch of the beverage could be distinctly different from the previous one, as the producer does not intervene in the fermentation process.
The result of this is twofold: firstly, these styles seen as mentally \say{refreshing}, meaning that they represent unique flavours not encountered in any commercial beers and secondly, the degree of acceptance of these unique flavours are incredibly difficult to quantify and predict.
The latter arises from the fact that there is now the per-batch variation, unique flavour combinations, uncommon structure of the beer - all at play when assigning the final score.
This makes interpretation of \textit{what makes a wild ale good} a very challenging task (as some flavours not originally intended by the producer may, in fact, improve the scores of a specific batch of beverages).

\subsection{Lack of standardisation}
Craft beer world is rich on novel styles, ingredients and flavour combinations, allowing the breweries to not be limited by any formal definitions during the production.
However, this also makes any sort of normalisation incredibly challenging: some producers give unique names to the styles they present (e.g. Vault City Brewing - Iron Brew as a beer style) which fall outside any naming conventions, even the ones as flexible as the ones used in craft beer brewing.
As a result, we had to construct a separate style category named \textit{Specialty and hybrid styles}, which include styles that cannot be aggregated into any other bucket.
Specifically the \say{barrel aged} (which was likely an old brown), the \say{iron brew} (which we could not come up with a style for) and the \say{basilikum-infused blond ale} (which technically is marketed as a blond ale, but includes ingredients not typically present in this style, making any comparison inherently unfair).
We suspect that this problem can be further exacerbated in the future, given the creativity of certain craft beer producers.
While it might be a relatively difficult problem to solve in the context of data analytics, the judges, nonetheless, welcome such experiments and have thoroughly enjoyed trying these novel style combinations.

\subsection{Communication cost}
Given the distributed nature of the protocol, it is important to discuss the implications of the communication cost on DBA.
In contrast to normal FL, DBA has two types of communication cost: the message broadcasting cost and the comprehension cost.
As the majority of the federation are considered to be polite (with one or two occasional exceptions), the broadcasted message may often involve words such as \say{could you} and \say{please}, inflating the cost of communication.
Over the tasting session, however, the politeness coefficient drops and the broadcasting becomes more abrupt and, in many case, starts missing syllables in order to optimise the total communication cost.
This is done in order to compensate for the opposite effect in the comprehension cost, which typically rises in parallel to the number of tasting rounds completed during a single instantiation of DBA.
As a result, these two types of communication costs change over time, but are (at least in theory) supposed to balance each other out even in lengthy DBA protocols which are struggling to converge.
While the overall communication time remained relatively consistent during our study due to this phenomenon, we cannot make an assumption that this definitely holds across all types of federations.
The investigation into the interplay of these two parts of the communication protocol is left as the future work.
\subsection{Naming conventions}
In this work we refer to all alcoholic (or otherwise) malted beverages as \say{beer}.
However, such naming conventions do not hold in every single country.
In particular, Germany, has came up with a special law called Reinheitsgebot (dated to 1516 \cite{reinheitsgebot1516}), which states that beer is a beverage that consists of only four ingredients (with a rare exception of the fifth - coriander): water, yeast, malt and hops.
The majority of data points of our study, sadly, do not fit this definition.
So for the researchers from Germany we need to clarify that while the conclusions apply to both \say{the real beer} (the one defined as such in Germany) and the malt-based alcoholic beverages, we do, in fact, abuse the notation in this study to make it more applicable to other, more permissive and forward-facing countries.

\subsection{Collaborative voting}
As evident from Section \ref{sec:results}, the there is a significant difference between the individual judge preferences and the final collaborative rankings.
Overall we note that on their own, judges have highly biased personal preferences (also as evident from their self-description given to LLMs as part of the experiments), meaning that collaborative aggregation helps reduce the individual bias in the results and provide a clearer picture of consumer preferences.
Therefore, we argue that aggregation of multiple diverse preferences and opinions is essential to derive actionable insights for such data: many styles which are \say{interesting} have drastically different perception based on the preferences of the end consumer.
A very visual example is the \say{Parfait Moment} sour ale, which had a very intense aromas of tropical fruits and vanilla, scoring $4.0$, $4.1$ and $2.5$ respectively, showcasing that unusual flavour combinations can have a very different reception based on the taster.

\section{Limitations and future work}
No study comes without limitations and potential for follow-up experiments, so we dedicate this section to talking about these in greater detail. 

\subsection{Optimal hyperparameter selection}
In this study we treated the number of judges vs the rest of the federation as an arbitrarily set hyperparameter.
It is entirely possible that under a different set up (with fewer leaders and more community effort on behalf of the federation), the results obtained could be much more reliable and represent additional trends not captured during this study.
We leave the exploration of this as part of future work, as authors are far too strongly opinionated regarding the leader selection and only have a small number of individuals who would willingly participate in such study. 

\subsection{Geographic limitations}
As the study was conducted in Vienna, this placed certain limitations both on the diversity of the datapoints as well as on the size (and composition) of the federation.
Since this is mostly a European event (including its former member - the United Kingdom), the beverages were more representative of the trends in craft beer consumption among the European population.
This means that while many of these trends may hold in other geographies, authors do not claim such outcome to be guaranteed.
We do note, however, that we consider the federation to be diverse in the context of this study: participants have vastly different background, nationalities and prior experiences in the world of craft beer consumption.
As a result, while there could have been certain regional biases when it comes to score assignments, we still consider the results to be representative.

\subsection{Scope of the study}
As all authors have their full-time responsibilities (being employed as well as having additional commitments), several parts of this work had to be left as future projects.
For instance, in the original iteration of the work, we planned to fine-tune a small language model (SLM) to be used as a personal beer companion and evaluate its effectiveness compared to the baseline models discussed in the experimental section.
The idea was to learn the trends and patterns discussed in this study as well as any additional similar studies (as informal as they may be) on services associated with sharing qualitative beer reviews (e.g. the Untappd project \cite{untappd}).
Additionally, after making a full recovery, the authors only had limited time outside of their working hours to contribute to this project before the deadline (which we set to be the $1$st of April, for no specific reason).
Unfortunately, such extensions require a fair amount of additional work, motivation and GPU budget that none of the authors had.
Therefore, we, sadly, had to leave the majority of complex ideas as future work. 
The authors would welcome any support from the wider beer appreciation community in order to continue this project.
This could ideally take shape of data generation, processing and sanitation, as review-sharing services are well-known to be difficult due to a) the quality of the reviews, b) the presence of large quantities of noise and c) the restrictions placed on the automated data collection.

\section{Conclusions}
In this work we present a framework to address the growing gap between the producers and the consumers of various types of craft beer.
We propose a  first-of-its-kind framework for collaborative beer review aggregation and analysis, evaluate it on a real-world dataset collected for the purposes of this study and share the insights that were gained using our framework.
Additionally, we empirically validate the need in such studies by comparing the trends discovered and consumption preferences obtained through our collaborative analysis against various popular LLMs.
We demonstrate that a) LLMs are not yet very familiar with recent developments in craft beer consumption trends and b) that they cannot (yet) be used as personalised beer recommendation systems.

On a personal note, we want to highlight that the science of this study should really be taken half-seriously.
We deliberately chose light-hearted language to wrap the core message of this paper: the world of craft beers is wonderfully diverse, complex and often underappreciated.
Many of craft beer producers have suffered a great deal during COVID-$19$ pandemic as well as due to being outcompeted by commercial producers, who often have the economies of scale to their advantage.
As a result, many wonderfully creative producers (our personal highlights include the authors' favourite microbrewery TrueBrew and the HopDog tap house) have struggled to stay afloat.
So we urge you to, please, support your local craft beer producers - without them the world of malted beverages would be significantly less exciting, a lot more rigid and lack really novel, experimental works which many of us would really enjoy.

\section*{Acknowledgements}
We would like to formally thank Dr. Grigory Tagiltsev, Dr. Jonathan Kaufman, Maria Ciapponi, Martina Cafiso and Eva Triantopoulou. Finally, we would like to extend our gratitude to the team of the Vienna Kraft brewery for hosting such an amazing internation tasting event. 
\bibliography{main}
\bibliographystyle{icml2021}
\newpage
\section{Appendix}
% \begin{figure*}[!h]
% \centering
\begin{minipage}{0.90\textwidth}

\begin{tcolorbox}[
  colback=gray!4,
  colframe=black,
  boxrule=0.6pt,
  arc=2mm,
  title={Prompt template used for all evaluated models},
  fonttitle=\bfseries,
  breakable
]
\begin{lstlisting}[style=promptstyle]
You are an expert beer sommelier.

You are given:
1) A list of beverages (with style, ABV, and other attributes if available)
2) Three consumer profiles

Your tasks:

---------------------
TASK 1: RECOMMENDATION
---------------------

For EACH consumer (A, B, C):

- Recommend EXACTLY 5 beverages from the provided list
- Do NOT recommend any beverage not present in the list
- Rank the recommendations from 1 (best match) to 5

For each recommendation, provide:
- beverage_name
- rank (1-5)
- justification (max 40 words)

Your recommendations must be based ONLY on:
- alignment with stated taste preferences
- style compatibility
- intensity (e.g. ABV, bitterness, sweetness)
- novelty vs familiarity (if stated in profile)

---------------------
TASK 2: TREND SUMMARY
---------------------

Provide a concise summary (max 120 words) of recent craft beer consumer trends.

Requirements:
- Focus ONLY on the trends you can derive from the data provided as input (e.g. many IPAs, fruited sours, etc.)
- Do NOT fabricate data, trends or statistics, only use what it given as the input.

---------------------
INPUT
---------------------

Profile A:
{...}

Profile B:
{...}

Profile C:
{...}

Beverage list:
{brewery, beer_name, beer_style, abv_percent, ...}
\end{lstlisting}
\end{tcolorbox}
% \caption{Abbreviated version of the prompt used in the experiment. The full consumer profiles and beverage catalogue are replaced with placeholders for readability.}
\end{minipage}
% \end{figure*}

% In the document
\begin{figure*}[t]
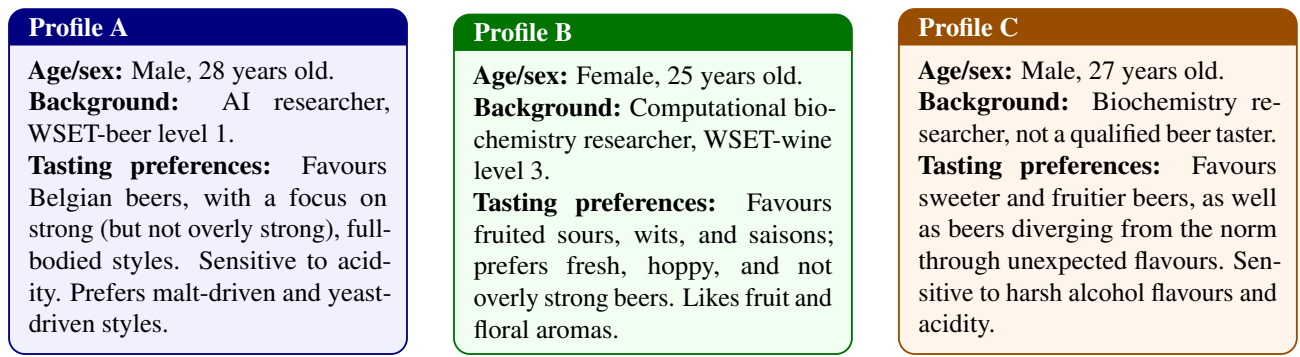

\centering

\begin{minipage}[t]{0.31\textwidth}
\begin{tcolorbox}[
  colback=blue!6,
  colframe=blue!50!black,
  boxrule=0.7pt,
  arc=2mm,
  title={Profile A},
  fonttitle=\bfseries,
  left=1.5mm,right=1.5mm,top=1mm,bottom=1mm
]
\textbf{Age/sex:} Male, 28 years old.\\
\textbf{Background:} AI researcher, WSET-beer level 1.\\
\textbf{Tasting preferences:} Favours Belgian beers, with a focus on strong (but not overly strong), full-bodied styles. Sensitive to acidity. Prefers malt-driven and yeast-driven styles.
\end{tcolorbox}
\end{minipage}
\hfill
\begin{minipage}[t]{0.31\textwidth}
\begin{tcolorbox}[
  colback=green!6,
  colframe=green!45!black,
  boxrule=0.7pt,
  arc=2mm,
  title={Profile B},
  fonttitle=\bfseries,
  left=1.5mm,right=1.5mm,top=1mm,bottom=1mm
]
\textbf{Age/sex:} Female, 25 years old.\\
\textbf{Background:} Computational biochemistry researcher, WSET-wine level 3.\\
\textbf{Tasting preferences:} Favours fruited sours, wits, and saisons; prefers fresh, hoppy, and not overly strong beers. Likes fruit and floral aromas.
\end{tcolorbox}
\end{minipage}
\hfill
\begin{minipage}[t]{0.31\textwidth}
\begin{tcolorbox}[
  colback=orange!8,
  colframe=orange!60!black,
  boxrule=0.7pt,
  arc=2mm,
  title={Profile C},
  fonttitle=\bfseries,
  left=1.5mm,right=1.5mm,top=1mm,bottom=1mm
]
\textbf{Age/sex:} Male, 27 years old.\\
\textbf{Background:} Biochemistry researcher, not a qualified beer taster.\\
\textbf{Tasting preferences:} Favours sweeter and fruitier beers, as well as beers diverging from the norm through unexpected flavours. Sensitive to harsh alcohol flavours and acidity.
\end{tcolorbox}
\end{minipage}

\caption{Consumer profiles used for prompting the language models in the recommendation experiment.}
\label{fig:beer_profiles}
\end{figure*}

\begin{figure*}
    \begin{tcolorbox}[
  enhanced,
  breakable,
  colback=gray!4,
  colframe=black,
  boxrule=0.7pt,
  arc=2mm,
  title={Model-predicted craft beer consumption trends},
  fonttitle=\bfseries,
  left=2mm,right=2mm,top=1.5mm,bottom=1.5mm
]

\textbf{Gemma 3 4B:} Recent craft beer trends include a strong emphasis on hazy IPAs, driven by the desire for juicy, textured beers. Fruited sours are also prominent, leveraging seasonal fruits for bright, refreshing styles. Low-ABV beers, particularly wheat beers and lagers, are gaining traction. There is also a noticeable increase in pastry stouts and IPAs with complex, layered flavours, reflecting growing consumer interest in premium, experimental beers.

\vspace{1mm}

\textbf{Llama 3.2 3B:} Recent craft beer trends include many hazy IPAs, growing popularity of low-ABV beers, fruited sours and pastry stouts due to their distinctive flavour profiles, and a broader rise in sour beers, including wild ales and lacto-fermented styles.

\vspace{1mm}

\textbf{Ministral 3B:} Recent craft beer trends in the provided data highlight a strong preference for fruited sours, Belgian-style beers, and hazy IPAs / hazy pale ales. Low-ABV beers dominate, reflecting a shift toward lighter and more approachable styles. Fruited and floral notes are prominent, while imperial and stout styles appear less common, suggesting a trend toward novelty and balance over intensity.

\vspace{1mm}

\textbf{Phi-4:} Recent craft beer consumer trends indicate a strong preference for fruited sours and IPAs, particularly hazy varieties. There is notable interest in low-ABV beers, with sour styles such as gose gaining traction. Unique and unexpected flavour combinations, including fruit-infused stouts and pastry-inspired brews, are also popular. More broadly, the output suggests a trend toward blending traditional beer styles with novel ingredients to create distinct sensory experiences.

\vspace{1mm}

\textbf{Qwen3 4B:} The model did not provide a usable trend summary. Instead, it returned profile-specific recommendation commentary and constraints related to the prompt, rather than aggregate observations about the beer list or broader consumer trends.

\vspace{1mm}

\textbf{Qwen2.5 3B:} Recent trends indicate a rise in hazy IPAs and low-ABV sours, reflecting preferences for fresher, fruitier, and lighter styles. The model also identifies continued interest in strong and malty stouts, as well as sour ales with unexpected flavour combinations.

\end{tcolorbox}
\caption{Trend summaries produced by the evaluated language models in response to the shared prompting setup. Qwen3 4B did not generate a valid trend summary and is therefore reported separately.}
\label{fig:llm_trend_predictions}
\end{figure*}

\begin{figure*}[h!]
    \centering
    \includegraphics[width=0.92\textwidth]{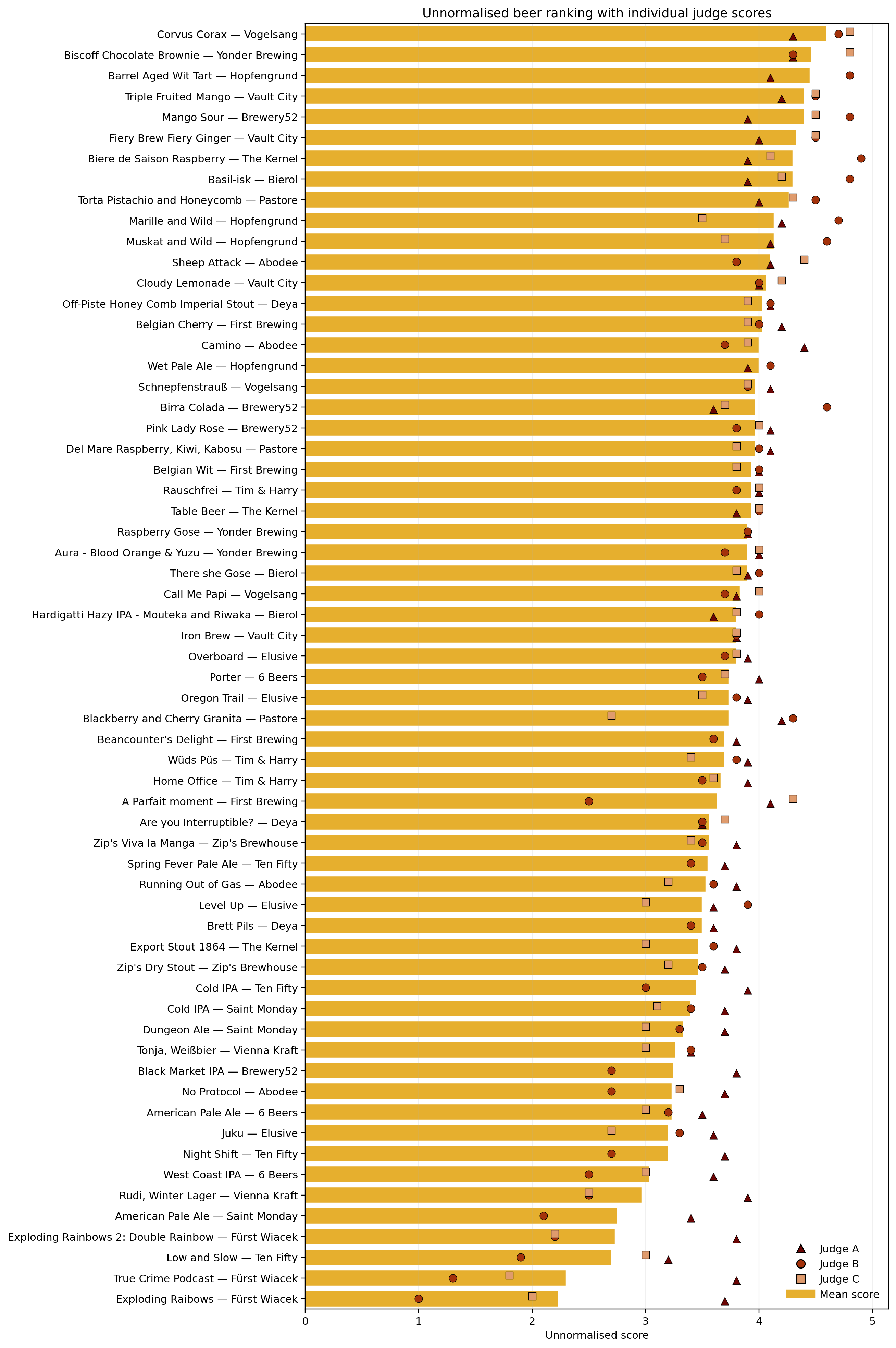}
    \caption{Numerical scores and ranking for each participating beverage, unnormalised.}
    \label{fig:complete_ranking}
\end{figure*}

\begin{figure*}[h!]
    \centering
    \includegraphics[width=\textwidth]{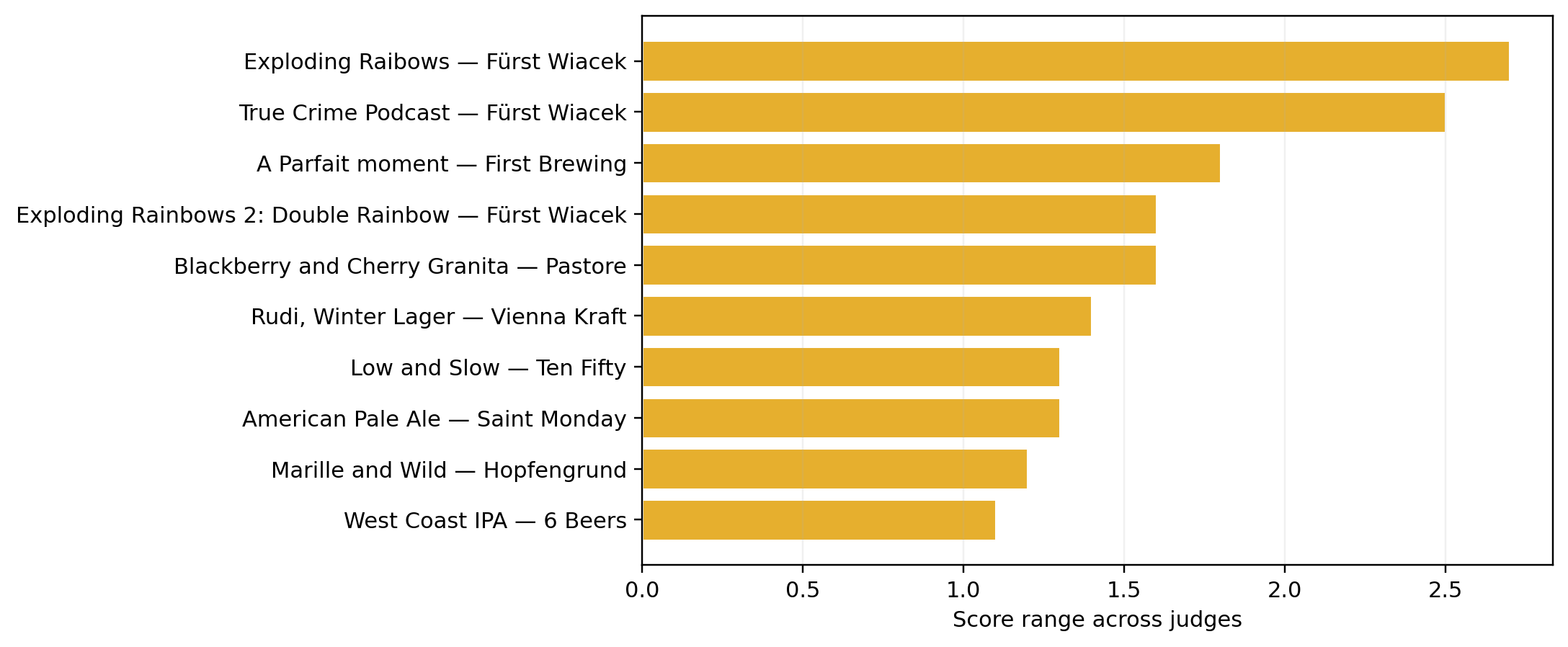}
    \caption{Beers with the largest score variations, unnormalised. Higher means less agreement between the judges.}
    \label{fig:divisive_beers}
\end{figure*}

\end{document}